# Bifurcation in the Growth of Continental Crust

Dennis Höning[a,b,c*], Nicola Tosi[c,d], Hendrik Hansen-Goos[e], Tilman Spohn[c]

[a] Origins Center, Nijenborgh 7, 9747 AG Groningen, The Netherlands
[b] Department of Earth- and Life Sciences, Vrije Universiteit Amsterdam, Amsterdam, The Netherlands
[c] Institute of Planetary Research, German Aerospace Center (DLR), Berlin, Germany
[d] Department of Astronomy and Astrophysics, Technische Universität Berlin, Berlin, Germany
[e] Institute for Theoretical Physics, Eberhard Karls Universität Tübingen, Tübingen, Germany
*Corresponding author. Email: d.hoening@vu.nl

ABSTRACT

Is the present-day water-land ratio a necessary outcome of the evolution of plate tectonic planets with a similar age, volume, mass, and total water inventory as the Earth? This would be the case – largely independent of initial conditions – if Earth's present-day continental volume were at a stable unique equilibrium with strong self-regulating mechanisms of continental growth steering the evolution to this state. In this paper, we question this conjecture. Instead we suggest that positive feedbacks in the plate tectonics model of continental production and erosion may dominate and show that such a model can explain the history of continental growth.

We investigate the main mechanisms that contribute to the growth of the volume of the continental crust. In particular, we analyze the effect of the oceanic plate speed, depending on the area and thickness of thermally insulating continents, on production and erosion mechanisms. Effects that cause larger continental production rates for larger values of continental volume are positive feedbacks. In contrast, negative feedbacks act to stabilize the continental volume. They are provided by the increase of the rate of surface erosion, subduction erosion, and crustal delamination with the continental volume. We systematically analyze the strengths of positive and negative feedback contributions to the growth of the continental crust. Although the strengths of some feedbacks depend on poorly known parameters, we conclude that a net predominance of positive feedbacks is plausible. We explore the effect of the combined feedback strength on the feasibility of modeling the observed small positive net continental growth rate over the past 2-3 billion years. We show that a model with dominating positive feedbacks can readily explain this observation in spite of the cooling of the Earth's mantle acting to reduce the continental production rate. In contrast, explaining this observation using a model with dominating negative feedbacks would require the continental erosion and production rates to both have the same or a sufficiently similar functional dependence on the thermal state of the mantle, which appears unreasonable considering erosion to be largely dominated by the surface relief and weathering.

The suggested scenario of dominating positive feedbacks implies that the present volume of the continental crust and its evolution are strongly determined by initial conditions. Therefore, exoplanets with Earth-like masses and total water inventories may substantially differ from the Earth with respect to their relative land/surface ratios and their habitability.

### 1. Introduction

Having both emerged land as well as oceans is an important environmental condition of the present-day Earth for sustaining its specific biosphere (e.g., Zahnle et al., 2007; Sleep, 2018). 40% of the Earth's surface is covered by continental crust of which 13% is continental shelf area under water. Emerged continents provide direct access to solar energy and nutrients. As a result, most of the bioactivity is found on continents and continental shelves, while the deep oceans are mostly desert (e.g., Behrenfeld and Falkowski, 1997; Kallmeyer et al., 2012). On the other hand, the ocean surface feeds the rain water cycle, such that a much smaller ocean surface area would result in reduced rainfall and a drier climate. Ultimately, Earth's climate is regulated through weathering of emerged continents (Kasting and Catling, 2003; Foley, 2015) and weathering of the seafloor (Coogan and Gillis, 2013; Coogan and Dosso, 2015; Krissansen-Totton and Catling, 2017), so both mechanisms are presumably important in keeping Earth habitable over long time scales.

Upon considering the habitability of other planets, one may be tempted to assume that Earth's potential exoplanetary sister would have a similarly balanced ocean-land fraction as the present day Earth. But is such a configuration a natural outcome of the evolution of an Earth-like planet with plate tectonics? Is the present-day volume of Earth's continental crust a necessary result of its age, mass, volume, and total water inventory, but only weakly dependent on initial conditions?

Continental crust differs from the basaltic oceanic crust by having a granitic composition. With an average thickness of 40 km, it is much thicker than the oceanic crust (7 km on average). Despite some mechanisms that recycle continental crust into the Earth's mantle, it still contains rock samples from the early Earth (see review by Rudnick and Gao, 2003). While oceanic crust is formed at mid-ocean ridges and hot spots by partial melting of mantle material, the formation of continental crust is more complex. In subduction zones, the dehydration of oceanic crust, sediments, and forearc serpentinites releases water, thereby reducing the melting temperature of the subducting oceanic crust and mantle material. Partial melt with an andesitic composition is generated and migrates upwards. The formation of granites requires additional re-melting processes. Although the exact processes are poorly understood, water has been argued to be crucial in the formation of continental crust (Campbell and Taylor, 1983).

The timing, rates and mechanisms of continental growth over Earth's history are matters of intense debate. While some studies favor a rapid early growth (Armstrong and Harmon, 1981; Armstrong, 1991), others argue for a pulse of continental growth in the late Archean (Taylor and McLennan 1985; 1995), or invoke a steady, more gradual growth over time (Dhuime et al., 2012 and Belousova et al. 2010). Peaks in the rate of continental production have mainly been attributed to episodic behavior of the non-linear character of dynamic processes associated with mantle convection (e.g., O'Neill et al., 2013), and interactions between continents during supercontinent cycles (e.g., Stern, 2011).

Most studies agree that the volume of continental crust has never been larger than today and that the present-day rate of net continental growth is small (see review by Cawood et al., 2013). A notable exception is the work by Fyfe (1978), who argues that continental crust might have covered larger parts of the surface in the past and that the mechanism of growth then was different from todays. Using present-day continental growth mechanisms and thermal evolution models of the Earth to satisfy the combined observations of a small net growth rate and a continental volume that never exceeded its present day value in the past is challenging. The reason is that most continental production and erosion mechanisms that can be accounted for in simple thermal evolution models result in a decreasing net growth rate with mantle cooling. Examples are the melt fraction decreasing with temperature as well as the Rayleigh number, a measure of the convection strength, suggesting plate speed and water subduction rate to decrease with time. An effect to the contrary is the proposed thickening of the oceanic crust with increasing mantle temperature, which would tend to resist subduction (Sleep and Windley, 1982) and cause sluggish plate tectonics (Korenaga, 2006) for mantle temperatures higher than at present. However, the general trend should be that mantle cooling results in a decreasing rate of continental crust production with time, at least for the present-day style of plate tectonics. In contrast, the erosion rate of continental crust should not decrease significantly as the mantle cools. The main mechanisms of continental crust erosion are surface erosion, subduction erosion, and crustal delamination (Stern, 2011). Surface erosion depends on the total area of emerged continents with probably only a small influence of the mantle temperature. On the one hand, orogenic activity, and in turn surface erosion rates, may decrease upon mantle cooling due to a reduction of the plate speed. But on the other hand, mantle cooling may lead to an increase of the continental freeboard (e.g., Flament et al., 2008), causing an increasing surface erosion rate. Lower crustal delamination depends on the area and thickness of continents and, depending on the mechanism accounted for, possibly also on the thermal state of the mantle. Subduction erosion certainly does depend on the plate speed. However, crustal delamination and subduction erosion only make up part of the total erosion rate. Altogether, the continental production rate depends on the mantle temperature to a larger extent than the total continental erosion rate does.

If the present-day continental volume were in a stable equilibrium between production and erosion, then the foregoing discussion suggests that the stable equilibrium point should move to smaller values of the continental volume with time. A larger continental volume would then be expected for part of the Earth's history, which cannot be reconciled with the geological record. Alternatively, the present-day continental gain and loss rates may be far from an equilibrium state, which would imply a non-negligible present-day net growth rate. In this paper, we propose an alternative solution, where the present-day continental volume is near an unstable equilibrium point.

In our previous work, we modeled the continental and mantle water cycles as coupled systems together with the thermal evolution of the Earth (Höning et al., 2014; Höning and Spohn, 2016). We showed how mantle temperature, mantle water concentration and continental coverage define a phase space, and we investigated effects of the biosphere on the phase space. These effects were based on the assumption that the biosphere enhances the rate of weathering and erosion of continental crust, thus increasing the rate at which water-carrying sediments will enter subduction zones. As this enhances the rates of continental production and mantle water regassing, self-reinforcing mechanisms (or positive feedbacks) were established. Within the phase plane spanned by mantle water concentration and continental coverage, we concentrated on steady-state curves of both variables (along which their rates of change were zero), and on intersections between both curves defining fixed points of the system. We showed that for weak positive feedbacks one (stable) fixed point exists while for strong positive feedbacks three fixed points exist of which the intermediate fixed point is unstable with respect to continental coverage. The two other (stable) fixed points where located at smaller and larger values of the continental coverage. Such a transition can be explained by a pitchfork-bifurcation, as sketched in Fig. 1 (described in Section 2).

In our calculations, we chose initial conditions using a Monte-Carlo method to show that trajectories exist that evolve near the unstable fixed point. These trajectories are characterized by an early rapid growth of continental crust followed by a roughly constant continental crust volume (Höning and Spohn, 2016). The time of rapid growth was found to depend on the rate of change of the mantle temperature. For models with rapid cooling, growth took place early, more than 4 Gyr b.p. For models with more moderate cooling, rapid growth was delayed to 2.5 Ga b.p. However, most combinations of initial conditions were found to evolve towards the stable fixed points at much smaller or larger values of continental coverage, not representative of the present-day Earth.

Since we found that the system was stable with respect to the mantle water concentration in all our models, we will keep this constant in the present paper, focusing only on the continental volume. We systematically quantify the strength of positive and negative feedback processes related to the growth of the continental crust for the present Earth. We consider surface erosion, subduction erosion, and crustal delamination collectively as negative feedback processes and the thermally insulating effect of continents on the mantle as the positive feedback process. In order to quantify the latter, we assume that the oceanic heat flux per unit area increases with the volume of insulating continents and that the plate speed increases accordingly, resulting in a larger water subduction rate.

Our analysis shows that positive feedbacks can actually control the growth of the Earth's continental crust and may have done so for the past billions of years. In addition, we quantitatively show that only in a system dominated by positive feedbacks it is possible to explain a small positive net growth rate of $0.6 - 0.9 \, km^3 yr^{-1}$ (e.g., Dhuime et al., 2017) for 2 billion years or longer. This would imply that other planets, in particular exoplanets, may be characterized by a continental volume very different from that of the Earth, no matter how similar to the Earth in terms of size, composition and interior structure they are.

We start by introducing the theory of stability analysis and bifurcation in Section 2 and show how to apply it to the growth of continental crust. In Section 3, we describe the feedback mechanisms we account for. If the combined strength of the individual feedbacks exceeds zero, a bifurcation occurs in the growth of continental crust (illustrated in Fig. 1). In such a case, multiple equilibrium points corresponding to different values of the steady-state continental volume exist for similar planet parameters. In Section 4, we show the implications of dominating positive and dominating negative feedbacks on the net growth rate in the Earth's history and on the sensitivity of continental growth to initial conditions. In Section 5 we discuss the validity of our model and Section 6 concludes this paper.

## 2. Stability Analyses and Bifurcation

Consider a simple dynamical system that involves only one dynamical variable $x = x(t)$, which we also refer to as the observable. The observable is assumed to follow the dynamic equation

$$\dot{x} = f(x) \, . \tag{1}$$

Stationary points are of paramount importance in order to characterize both the dynamic and equilibrium properties of the system at hand. The stationary condition $\dot{x} = 0$ implies that we should study zeros of the function $f$. Assume that $x_0$ is a zero of $f$ and therefore a stationary point of the dynamical system. In order to study the behavior of the system near the stationary point, it is useful to expand $f$ in a Taylor series leading to $f(x) \approx f(x_0) + f'(x_0)(x - x_0) = f'(x_0)(x - x_0)$ since $f(x_0) = 0$ due to the stationary condition. Introducing $\Delta x(t) = x(t) - x_0$ and $\tau^{-1} = f'(x_0)$, Eq. (1) becomes

$$\dot{\Delta x} = \Delta x \, \tau^{-1} \, , \tag{2}$$

which can be readily integrated to give

$$\Delta x(t) = \Delta x(t_0) e^{(t-t_0)\tau^{-1}}, \tag{3}$$

where $\Delta x(t_0)$ denotes the distance from $x_0$ at time $t = t_0$.

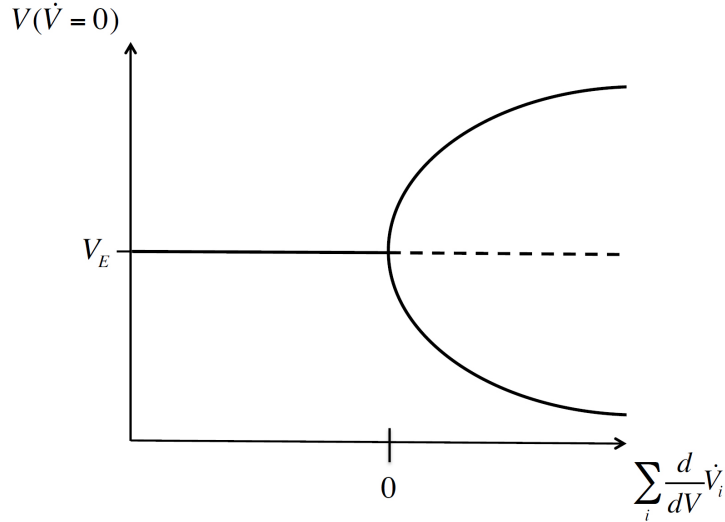

**Fig. 1:** Schematic diagram of a pitchfork-bifurcation representing the steady-state continental volume $V$ as a function of the combined strengths of the feedbacks in continental growth $\sum_i \frac{dV_i}{dV}$. It is assumed that the present-day continental volume, $V_E$, represents a fixed point. The solid and the dashed lines combine stable and unstable fixed points, respectively. If $\sum_i \frac{dV_i}{dV} < 0$, negative (stabilizing) feedbacks dominate, and only one fixed point exists. This fixed point acts as an attractor to the continental evolution, which would approach this fixed point. With increasing strength of the positive feedbacks, the system becomes less stable until at $\sum_i \frac{dV_i}{dV} = 0$ a bifurcation occurs. For $\sum_i \frac{dV_i}{dV} > 0$, the present-day continental volume is unstable, and two stable fixed points emerge at smaller and larger continental volume.

This simple result, valid only in close proximity of $x_0$, is the basis of stability analysis. It is clear that the sign of $\tau$ is significant for the system's behavior near $x_0$. If $\tau < 0$, i.e. $f'(x_0) < 0$, the stationary point is stable and a small deviation from $x_0$ decays exponentially with a life time $|\tau|$. The case $\tau > 0$, i.e. $f'(x_0) > 0$, corresponds to an unstable stationary point. A small deviation from $x_0$, be it positive or negative, grows exponentially. Clearly, a realistic physical system of finite size cannot allow for unlimited runaway exponential growth. Therefore, the system is bound to evolve towards a stable fixed point $x_+ > x_0$ for perturbations with $\Delta x > 0$ or into $x_- < x_0$ for $\Delta x < 0$.

In response to some external control parameter, a previously stable fixed point can become unstable. The stability analysis laid out above then leads to the phenomenon of bifurcation. According to random fluctuations the system either evolves toward $x_+$ or $x_-$. An example is the convective velocity at a given point in a Rayleigh-Bénard system when convection is initiated due to the Rayleigh number exceeding a critical value. The convective roll then receives its orientation as a result of random fluctuations.

In this paper we will apply the concept of stability analysis to the volume of continental crust $V$ as the dynamic variable, i.e. $x = V$. We are aware that the growth of continental crust is a complex process and that condensing it into a single variable, i.e. writing $\dot{V} = f(V)$, is a strong simplification. However, the basic concepts of linear stability analysis and bifurcation theory can be still applied provided that a few simplifying assumptions with respect to the real Earth system are introduced. In particular, we study contributions to $f$ only in the immediate vicinity of the state of the present day Earth, which – as will be discussed below – we consider as being at a stationary point.

An important feature of continental growth compared to systems typically investigated by stability analysis is its large life-time $\tau$; the growth of continental crust occurs over geological time scales. This opens the possibility that the Earth's continental growth system may remain near an unstable fixed point for millions or billions of years, up to the present day. This in turn would imply that the growth of the continental crust on planets in general could evolve differently and may strongly depend on initial conditions (e.g., on the initial mantle temperature or early forms of tectonics and crust growth as has recently been discussed for the Archean) and possibly on fluctuations that could drive the system towards one of the two stable branches with either low or high continental volume (see also Höning and Spohn, 2016).

### 3. Positive and Negative Feedbacks in Continental Growth

In this section we analyze the strengths of the positive and negative feedbacks in the growth of continental crust to address the question of whether the present-day Earth is located at the left side (dominating negative feedbacks) or right side (dominating positive feedbacks) of the bifurcation diagram (Fig. 1). The strengths of the feedbacks may strongly differ for different values of the continental volume and interior temperature (or heat flow) and we restrict our analysis to the present-day Earth, where we assume equilibrium between the continental production rate and the continental erosion rate. This assumption is a necessary requirement for the stability analysis. Both models of early continental growth (e.g., Armstrong and Harmon, 1981) and models of delayed continental growth (e.g., Taylor and McLennan, 1985) show a small, if any, net growth rate at present. Even models that predict a more gradual growth (Belousova et al., 2010; Dhuime et al., 2012) have small present-day net growth rates in common. Therefore, we assume that for the present day, $\dot{V} = \sum_i \dot{V}_i = 0$ where $\dot{V}_i$ are separate contributions to the net growth rate $\dot{V}$ of the continental crust volume $V$. Below, we will derive the function $f(V)$ such that $\dot{V} = f(V)$ in the vicinity of its present-day value. As outlined in Section 2, the present-day value of continental volume represents an unstable fixed point if $f'(V) = \frac{d\dot{V}}{dV} > 0$, i.e. if

$$\sum_i \frac{d\dot{V}_i^*}{dV} > 0 \ , \tag{4}$$

where the asterisk – here and in the analysis that follows – denotes non-dimensional values, scaled to the present-day Earth. If this inequality is satisfied, small changes in the volume of continental crust will tend to be amplified, and in the bifurcation diagram shown in Fig. 1 the present-day Earth will be located on the dashed line on the right side. In the following subsections, we discuss the relative strength of the various contributing feedback mechanisms $\frac{d\dot{V}_i^*}{dV}$.

Erosion mechanisms act as negative feedbacks. In Section 3.1, we consider the main mechanisms of continental erosion, namely surface erosion, subduction erosion, and crustal delamination (Stern, 2011). Continental production is directly related to the rate at which water is subducted and therefore to the plate speed. The relation between continental volume and plate speed is derived in Section 3.2, and in Section 3.3 we evaluate the individual feedback strengths.

### 3.1 Continental erosion and production

The main loss mechanisms of continental crust are subduction erosion, surface erosion, and crustal delamination (e.g., Stern, 2011). Feedbacks are established if these rates depend on the continental volume itself.

| Parameter | Value | Description | Equation |
|---|---|---|---|
| $R_E$ | $6.371 \cdot 10^6 \ m$ | Radius of the Earth | Eq. 6 |
| $\phi_{surf}$ | 1/3 | Present-day surface erosion rate relative to the total erosion rate | Eq. 8 |
| $\phi_{subd}$ | 1/3 | Present-day subduction erosion rate relative to the total erosion rate | Eq. 8 |
| $\phi_{dela}$ | 1/5 | Present-day crustal delamination rate relative to the total erosion rate | Eq. 8 |
| $n$ | 7 | Number of continents | Eq. 6 |
| $\alpha = \frac{A_{c,E}}{A_E}$ | 0.4 | Present-day surface fraction of continental crust | Eq. 6 |
| $\dot{V}_{prod,E}$ | $5.25 \ km^3/yr$ | Present-day continental production rate | Eq. 35 [1] |
| $V_E$ | $7.18 \cdot 10^9 \ km^3$ | Present-day continental volume | Eq. 35 [2] |
| $\chi = \frac{q_{mc,E}}{q_{mo,E}}$ | 0 to 0.42 | Present-day ratio between subcontinental and suboceanic heat flow | Eq. 32 |
| $\xi = \frac{\delta_{o,E}}{d_{c,E}}$ | 3 | Present-day ratio between the oceanic boundary layer thickness and the continental thickness | Eq. 32 |

**Tab. 1:** Constants and present-day Earth ratios used in the Model. [1] From Stern, 2011 [2] from Schubert and Sandwell, 1989

We start by discussing the feedback associated with surface erosion, which is responsible for $\approx 1/3$ of the total loss of continental crust (Stern, 2011). Scaling surface erosion is difficult, since it is a complex process dependent on factors such as the climate and topography (Bishop, 2007), which are not directly included in our model. As the upper layer of the surface of continental crust is eroded, it is reasonable to scale the surface erosion rate with the total surface area of continental crust. However, it is known that mechanical erosion increases with the topographic expression, which is a result of tectonic – particularly orogenic – activity, ultimately dependent on the plate speed. In contrast, chemical erosion is rather coupled to rainfall and not directly to topography. For our reference model, we neglect the effect of the plate speed on the surface erosion rate and set the rate of surface erosion $\dot{V}_{surf}$ proportional to the total surface area of continental crust $A_c$ only, which directly establishes a strong negative feedback. We discuss the effect of including the plate speed in the scaling of surface erosion in Section 5.5.

Formally, we have

$$\frac{\dot{V}_{surf}}{\dot{V}_{eros,E}} = \phi_{surf} A_c^* \ , \tag{5}$$

where $\dot{V}_{surf}$ is the rate of surface erosion, $\phi_{surf} = 1/3$, and $\dot{V}_{eros,E}$ is the total present-day continental erosion rate. The values of the different $\phi$ (and other parameter values) are summarized in Tab. 1. In this paper, we use the index $E$ to denote the present-day Earth value, i.e. $A_c^* = \frac{A_c}{A_{c,E}}$ where $A_c^*$ is the continental area relative to its present-day Earth value.

Similar to surface erosion, subduction erosion is also responsible for 1/3 of the total crustal loss (Stern, 2011), i.e. $\phi_{subd} = 1/3$. Subduction erosion is proportional to the plate speed $v$ and to the total length of subduction zones $L_s$. We will derive a relationship between the continental volume and the plate speed in the next section and come back to this feedback in Section 3.3. We set the total length of ocean-continent subduction zones proportional to the length of continental margins $L_c$. For relating the total length of continental margins to the total area of continental crust, $A_c$, we use stochastic geometry assuming continents as spherical caps as follows (see Höning et al., 2014)

$$\frac{L_c}{R_E} = 4\pi n (1 - \alpha A_c^*) \sqrt{\frac{1}{n} \ln[(1 - \alpha A_c^*)^{-1}] - \left(\frac{1}{n} \ln[(1 - \alpha A_c^*)^{-1}]\right)^2}, \qquad (6)$$

where $A_E$ is the Earth's total surface area, $R_E$ is the Earth's radius, $n$ the number of continents, and $\alpha = \frac{A_{c,E}}{A_E}$ is the present-day fraction of the Earth's surface covered by continental crust. For $\alpha = 0.4$ and $n = 7$, we get $\frac{L_{c,E}}{R_E} \approx 13.73$. Setting the total length of subduction zones proportional to $L_c$, i.e. $L_s^* = \frac{L_c}{L_{c,E}}$, we calculate $\frac{dL_s^*}{dA_c^*}$ for $n = 7$ continents:

$$\frac{dL_s^*}{dA_c^*} = \frac{R_E}{L_{c,E}} \frac{4\pi (2\ln^2(1 - 0.4A^*) + 16\ln(1 - 0.4A^*) + 7)}{5\sqrt{-\ln(1 - 0.4A^*)(\ln(1 - 0.4A^*) + 7)}} \qquad (7)$$

For $A_c^* = 1$, we get $\frac{dL_s^*}{dA_c^*} \approx -0.066$. Note that this value is smaller than zero, implying that the total length of continental margins will decrease for increasing continental area beyond the present value. However, continent-continent collision zones will increase at the same time, which counteracts the decreasing subduction erosion rate. Furthermore, the value of $\left|\frac{dL_s^*}{dA_c^*}\right|$ is very small for the present-day continental area. The total length of continental margins is close to the maximum here (which would be achieved at $\frac{A_c}{A_E} \approx 0.37$). Therefore, it is justified to neglect effects of the total length of subduction zones on continental erosion and production rates in the vicinity of the present-day Earth's continental area. Therefore, in determining the present-day feedback strengths, we will set $L_s^* = 1$.

If the lower continental crust is denser then the mantle asthenosphere, it is gravitational unstable and may sink into the mantle, a process known as crustal delamination (Bird, 1979). Delamination of the lower continental crust is responsible today for $\approx 1/5$ of the total crustal loss (Stern, 2011), i.e. $\phi_{dela} = 1/5$. Although continental delamination is a complex process, capturing the main parameters affecting delamination is crucial in order to describe the feedback accurately.

Several delamination mechanisms have been proposed. On the one hand, densification of the lowermost continental crust may be caused by metamorphic eclogitization (Krystopowicz and Currie, 2013). Since this phase transformation requires large pressures that may be caused by tectonic activity, the plate speed should play a role. Furthermore, the rate of crustal delamination should increase with the continental area and thickness since the continental volume that is gravitationally unstable would increase accordingly.

An important consequence of crustal delamination caused by phase changes is magmatism (Kay and Kay, 1993). Magmatism caused by crustal delamination could then also contribute to the formation of new continental crust. It has been argued that this mechanism has been important for the production of continental crust during the Archean (Zegers and van Keken, 2001). If delamination magmatism contributes to produce continental crust, an additional positive feedback would emerge.

On the other hand, Morency et al. (2002) describes an equilibrium thickness of the continental lithosphere at which the conductive geotherm crosses the mantle temperature. Due to thermal contraction, the lower part of the continental lithosphere can then become gravitationally unstable. It may be eroded either by small-scale convection at the base of the lithosphere or by edge-driven convection (Morency et al., 2002). This mechanism is not related to magmatism (Elkins-Tanton, 2005), and therefore represents a purely negative feedback. Besides depending on the continental area, the erosion rate also depends on the continental thickness, since isostasy causes the lower continental crust to sink deeper into the mantle as the continental thickness increases. It is a progressive process and not directly coupled to the plate speed.

Altogether, for our reference model, we treat delamination as a purely erosion mechanism of continental crust and set the rate of crustal delamination proportional to the continental area $A_c$ and continental thickness $z_c$ only. However, since plate speed may also play a role in particular for delamination by densification through phase changes, we also discuss the effect of scaling crustal delamination with continental area, continental thickness, and plate speed (Section 5.5).

Continental production is a complex process associated with the water transport in subduction zones. Although depending on details of the subduction zones, such as mantle temperature, the angle of subduction, and subduction rate of sediments, continental production should to first order be proportional to the plate speed. Note that the validity of this assumption is restricted to continental area not much larger than the present-day one, otherwise overlapping continents dominate for large values of continental coverage, and we would additionally have to account for the total length of subduction zones. However, for the Earth's history and the present-day, the continental volume did not exceed its present-day value, so we neglect the dependence of continental production on the total length of subduction zones.

Altogether, the net continental growth rate $\dot{V}_{net}$ for continental coverage not exceeding the present-day value can be written as

$$\dot{V}_{net} = \dot{V}_{prod,E} v^* + \dot{V}_{eros,E} \left( \phi_{surf} A_c^* + \phi_{subd} v^* L_s^* + \phi_{dela} A_c^* z_c^* + \phi_{rest} \right), \qquad (8)$$

with $\phi_{rest} = 1 - \phi_{surf} - \phi_{subd} - \phi_{dela}$, where $v^*$ is the plate speed relative to its present-day value. We assume that at the present-day, the net growth rate is zero:

$$\dot{V}_{prod,E} + \dot{V}_{eros,E} = 0 \ , \qquad (9)$$

which gives

$$\frac{\dot{V}_{net}}{\dot{V}_{prod,E}} = (v^* - 1) - \phi_{surf}(A_c^* - 1) - \phi_{subd}(v^* L_s^* - 1) - \phi_{dela}(A_c^* z_c^* - 1) \ . \quad (10)$$

As discussed earlier, we take $L_s^* = 1$ in the vicinity of the present-day value. Differentiating Eq. (10) with respect to the relative continental volume $V^*$ yields the feedback strength

$$\frac{d}{dV^*}\left( \frac{\dot{V}_{net}}{\dot{V}_{prod,E}} \right) = (1 - \phi_{subd}) \frac{dv^*}{dV^*} - \left( \phi_{surf} + \phi_{dela} \right) \frac{dA_c^*}{dV^*} - \phi_{dela} \frac{dz_c^*}{dV^*} \ . \quad (11)$$

If $\frac{d}{dV^*}\left( \frac{\dot{V}_{net}}{\dot{V}_{prod,E}} \right) > 0$, the positive feedback dominates. In the following section, we will derive an expression for $\frac{dv^*}{dV^*}$ to address this issue.

### 3.2 Continental volume and plate speed

Since continents act as insulators on the mantle heat, Earth's total mantle heat flow mainly passes through the oceanic lithosphere. Estimates for the present-day Earth suggest that the fraction of mantle heat flow passing through the continental surface area is only about 15% of the total (Jaupart et al., 2016). Therefore, it is reasonable to assume that the mantle heat flow through the oceanic crust per unit area increases with the continental volume.

We assume that the total mantle heat flow does not depend on the continental volume, as indicated by steady-state results of numerical simulations and laboratory experiments (Lenardic et al., 2005). Certainly, it becomes invalid for continents almost covering the entire planet. However, this paper focuses on the stability of the continental volume around its present-day value and on continental growth for Earth's recent history. Furthermore, insulating continents might reduce the global mantle cooling rate, which we discuss in Section 5.1. Setting the plate speed proportional to the convective flow rate is certainly a simplification and may be valid only as long as the tectonic mode does not change (discussed by Lenardic et al., 2005).

We start by deriving a value for $\frac{dv^*}{dA_c^*}$ assuming that continents act as perfect insulators (Section 3.2.1). In this case, the heat flow through the continental crust is zero and varying the continental thickness does not affect the plate speed, i.e. $\frac{dv^*}{dz_c^*} = 0$. In Section 3.2.2, we extend the model to also account for mantle heat flow into the continental lithosphere, which yields values for both, $\frac{dv^*}{dA_c^*}$ and $\frac{dv^*}{dz_c^*}$.

### 3.2.1 Continental crust as a perfect insulator

The total mantle heat flow $Q_m$ is the sum of the heat flow into the oceanic and continental lithosphere, which gives

$$Q_m = q_{mo}(A_E - A_c) + q_{mc}A_c , \qquad (12)$$

where $q_{mo}$ and $q_{mc}$ are the mantle fluxes per unit area into the oceanic and continental lithosphere, respectively. As discussed above, we assume that the total mantle heat flow $Q_m$ is not affected by the continental volume, setting $Q_m = Q_{m,E}$, which gives

$$q_{mo}(A_E - A_c) + q_{mc}A_c = q_{mo,E}(A_E - A_{c,E}) + q_{mc,E}A_{c,E} . \qquad (13)$$

We write again the relative values with respect to the present-day Earth with an asterisk, i.e. $A^* = \frac{A_c}{A_{c,E}}$, $q_{mo}^* = \frac{q_{mo}}{q_{mo,E}}$, and $q_{mc}^* = \frac{q_{mc}}{q_{mc,E}}$. We further introduce $\chi = \frac{q_{mc,E}}{q_{mo,E}}$, describing the ratio between the present-day basal continental and oceanic heat fluxes, and again use $\alpha = \frac{A_{c,E}}{A_E}$. From Eq. (13), we get

$$q_{mo}^*(1 - \alpha A^*) + q_{mc}^*\alpha\chi A^* = (1 - \alpha) + \alpha\chi . \qquad (14)$$

For $\chi = 0$, which is assumed in this section for simplicity, we have

$$q_{mo}^* = \frac{1 - \alpha}{1 - \alpha A^*} . \qquad (15)$$

Following boundary layer theory, the convection rate is a function of the Rayleigh number $Ra$, which in turn scales with the heat flux. Setting the plate speed $v$ proportional to the convection rate, we have

$$v^* = Ra^{*(2\beta)} = q_{mo}^{*2} = \left(\frac{1-\alpha}{1-\alpha A^*}\right)^2 .$$  (16)

Differentiating with respect to the continental area and setting $A^* = 1$ and $\alpha = 0.4$ yields

$$\frac{dv^*}{dA^*} = 2 \cdot 0.4 \frac{(1-0.4)^2}{(1-0.4A^*)^3} = \frac{4}{3} .$$  (17)

### 3.2.2 Mantle heat flux and continental thickness

We start by calculating the temperature beneath the continental crust $T_c$ using the steady-state continental geotherm (e.g., Turcotte and Schubert, 2014)

$$T_c = T_s + q_{mc}\frac{d_c}{k} + c_1\frac{d_c^2}{k} ,$$  (18)

where $T_s$ is the surface temperature, $k$ is the thermal conductivity (for simplicity, we take the same $k$ for the continental crust and the lithospheric mantle), and $c_1$ is a constant representing the heat production within the continental crust. We use boundary layer theory to relate the subcontinental temperature to the mantle temperature $T_m$

$$T_c = T_m - q_{mc}\frac{\delta_c}{k} ,$$  (19)

where $\delta_c$ is the thickness of the subcontinental boundary layer. For the subcontinental mantle heat flux, we have

$$q_{mc} = \frac{k(T_m-T_s)-c_1 d_c^2}{d_c+\delta_c} .$$  (20)

Keeping the mantle and surface temperatures fixed, the subcontinental mantle heat flux of the present-day Earth, $q_{mc,E}$, is

$$q_{mc,E} = \frac{k(T_m-T_s)-c_1 d_{c,E}^2}{d_{c,E}+\delta_{c,E}} ,$$  (21)

which gives for the constant $c_1$

$$c_1 = \frac{k(T_m-T_s)-q_{mc,E}(d_{c,E}+\delta_{c,E})}{d_{c,E}^2} .$$  (22)

Combining Eqs. (20) and (22), we have

$$q_{mc}(d_c + \delta_c) = k(T_m - T_s) - \frac{d_c^2}{d_{c,E}^2}\left(k(T_m - T_s) - q_{mc,E}\left(d_{c,E} + \delta_{c,E}\right)\right) .$$  (23)

We now introduce the nondimensional parameters $q_{mc}^* = \frac{q_{mc}}{q_{mc,E}}$, $q_{mo}^* = \frac{q_{mo}}{q_{mo,E}}$, $d_c^* = \frac{d_c}{d_{c,E}}$ and use the heat flux through the present-day oceanic lithosphere

$$q_{mo,E} = \frac{k(T_m - T_s)}{\delta_{o,E}},$$ (24)

with the thickness of the present-day oceanic boundary layer $\delta_{o,E}$, to get from Eq. (23)

$$q_{mc}^* q_{mc,E}(d_c^* d_{c,E} + \delta_c) = q_{mo,E}\delta_{o,E} - d_c^{*2}\left(q_{mo,E}\delta_{o,E} - q_{mc,E}\left(d_{c,E} + \delta_{c,E}\right)\right).$$ (25)

We follow boundary layer theory and set the Rayleigh number at the bottom of the sub-oceanic thermal boundary layer equal to the Rayleigh number of the sub-continental thermal boundary layer:

$$\frac{\alpha c_p \rho^2 g\,(T_m - T_s)\delta_o^3}{\eta k} = \frac{\alpha c_p \rho^2 g\,(T_m - T_c)\delta_c^3}{\eta k}$$ (26)

Note that the material parameters (thermal expansion coefficient $\alpha$, specific heat capacity $c_p$, gravitational acceleration $g$, viscosity $\eta$, and thermal conductivity $k$) are constant as they refer to the same mantle temperature $T_m$, which we keep constant. Using $q_{mc} = \frac{k}{\delta_c}(T_m - T_c)$ and $q_{mo} = \frac{k}{\delta_o}(T_m - T_s)$, we can write Eq. (26) as

$$\delta_c = \delta_o \left(\frac{q_{mo}}{q_{mc}}\right)^{\frac{1}{4}}$$ (27)

and

$$\delta_{c,E} = \delta_{o,E} \left(\frac{q_{mo,E}}{q_{mc,E}}\right)^{\frac{1}{4}}.$$ (28)

The relative sub-oceanic boundary layer thickness depending on the heat flux is

$$\frac{\delta_o}{\delta_{o,E}} = \frac{q_{mo,E}}{q_{mo}}.$$ (29)

Combining Eqs. (25) and (27) to (29), we get

$$q_{mc}^* d_c^* q_{mc,E} d_{c,E} + \frac{q_{mc}^*}{q_{mo}^*} q_{mc,E}\delta_{o,E}\left(\frac{q_{mo}^*}{q_{mc}^*}\right)^{\frac{1}{4}}\left(\frac{q_{mo,E}}{q_{mc,E}}\right)^{\frac{1}{4}}$$
$$= q_{mo,E}\delta_{o,E} - d_c^{*2}\left(q_{mo,E}\delta_{o,E} - q_{mc,E}\left(d_{c,E} + \delta_{o,E}\left(\frac{q_{mo,E}}{q_{mc,E}}\right)^{\frac{1}{4}}\right)\right)$$ (30)

We now divide Eq. (30) by $q_{mc,E} d_{c,E}$ to obtain

$$q_{mc}^* d_c^* + \left(\frac{q_{mc}^*}{q_{mo}^*}\right)^{\frac{3}{4}}\frac{\delta_{o,E}}{d_{c,E}}\left(\frac{q_{mo,E}}{q_{mc,E}}\right)^{\frac{1}{4}} = \frac{q_{mo,E}}{q_{mc,E}}\frac{\delta_{o,E}}{d_{c,E}} - d_c^{*2}\left(\frac{q_{mo,E}}{q_{mc,E}}\frac{\delta_{o,E}}{d_{c,E}} - 1 - \frac{\delta_{o,E}}{d_{c,E}}\left(\frac{q_{mo,E}}{q_{mc,E}}\right)^{\frac{1}{4}}\right)$$
(31)

We introduce the present-day constants $\xi = \frac{\delta_{o,E}}{d_{c,E}}$ and $\chi = \frac{q_{mc,E}}{q_{mo,E}}$ and get

$$q_{mc}^* d_c^* + \left(\frac{q_{mc}^*}{q_{mo}^*}\right)^{\frac{3}{4}} \xi \chi^{-\frac{1}{4}} = \xi \chi^{-1} - d_c^{*2}\left(\xi \chi^{-1} - 1 - \xi \chi^{-\frac{1}{4}}\right). \tag{32}$$

Combining Eqs. (14) and (32) allows us to relate the plate speed to the continental area and thickness. In solving for $\frac{dv^*}{dA_c^*}$ and $\frac{dv^*}{dz_c^*}$ numerically, we use $\xi = 3$, which is a reasonable value when assuming $\delta_{o,E} \approx 100\ km$ and $z_c^* \approx 35\ km$. The value for $\chi$ is less certain, because the heat production within the continental crust relative to the continental surface heat flow is not well known. If all continental heat flow is caused by radioactive decay within the continental crust, we have $\chi = 0$, reproducing our previous result of $\frac{dv^*}{dA_c^*} = \frac{4}{3}$ and $\frac{dv^*}{dz_c^*} = 0$ (see Section 3.2.1). In contrast, a maximum value of $\chi = 0.42$ is used as an upper bound following Korenaga (2008), who combined the continental heat flow of Jaupart et al. (2007) with the lowest possible heat production of Rudnick and Gao (2003), weighted by the relative continental and oceanic areas. As a mean value, we also test $\chi = 0.2$. In Fig. 2a, we plot the relative plate speed $v^*$, as a function of the relative continental area $A_c^*$ (blue) and relative continental thickness $z_c^*$ (red), keeping the continental thickness (and area, respectively) constant. In Fig. 2b, we plot the respective derivatives $\frac{dv^*}{dA_c^*}$ (blue) and $\frac{dv^*}{dz_c^*}$ (red). The dotted lines refer to $\chi = 0$, the solid lines to $\chi = 0.2$, and the dashed lines to $\chi = 0.42$. In the vicinity of the present-day continental area and thickness, for $\chi = 0.2$ we get $\frac{dv^*}{dA_c^*} = 0.97$ and $\frac{dv^*}{dz_c^*} = 1.11$, and for $\chi = 0.42$ we get $\frac{dv^*}{dA_c^*} = 0.64$ and $\frac{dv^*}{dz_c^*} = 0.71$.

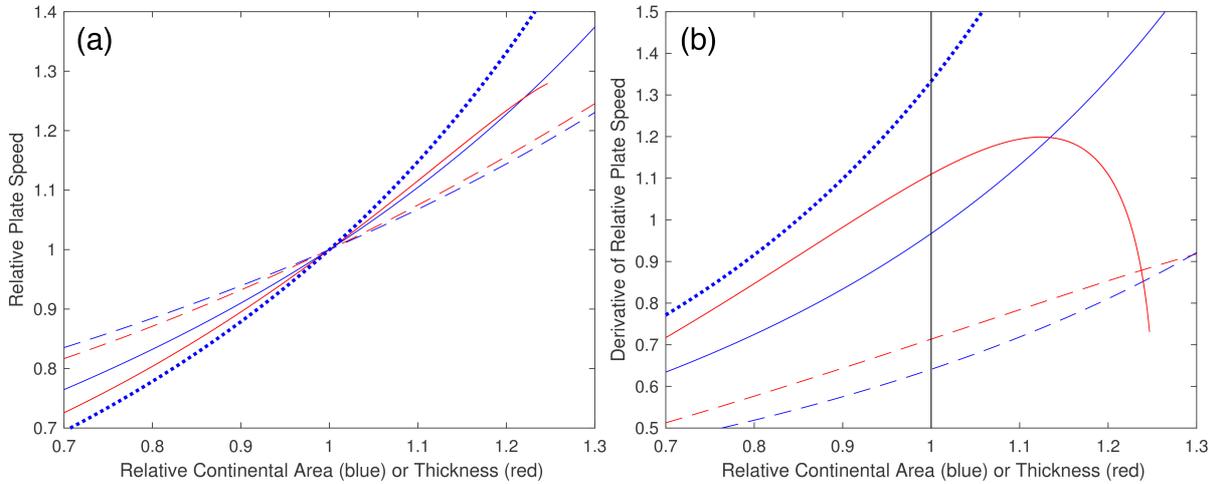

**Fig. 2:** Relative plate speed (a) and its derivative (b) as functions of the relative continental area keeping the thickness constant (blue) and as functions of the relative continental thickness keeping the area constant (red). The thick dotted curve applies to the model of perfectly insulating continents, i.e. $\chi = 0$, the solid curves to $\chi = 0.2$, and the dashed curves to $\chi = 0.42$.

### 3.3 Evaluation of the strengths of the feedbacks

Using the values for $\frac{dv^*}{dA_c^*}$ and $\frac{dv^*}{dz_c^*}$, we write Eq. (11) as

$$\frac{d}{dV^*}\left(\frac{\dot{V}_{net}}{\dot{V}_{prod,E}}\right) = (1 - \phi_{subd})\frac{dv^*}{dA_c^*}\frac{dA_c^*}{dV^*} + (1 - \phi_{subd})\frac{dv^*}{dz_c^*}\frac{dz_c^*}{dV^*} - \left(\phi_{surf} + \phi_{dela}\right)\frac{dA_c^*}{dV^*} - \phi_{dela}\frac{dz_c^*}{dV^*}$$

<div align="right">(33)</div>

Using the parameter values $\frac{dv^*}{dA_c^*} = 0.97$ and $\frac{dv^*}{dz_c^*} = 1.11$ (as derived for $\chi = 0.2$), $\phi_{surf} = \phi_{subd} = \frac{1}{3}$ and $\phi_{dela} = \frac{1}{5}$, we have

$$\frac{d}{dV^*}\left(\frac{\dot{V}_{net}}{\dot{V}_{prod,E}}\right) = 0.11\frac{dA_c^*}{dV^*} + 0.54\frac{dz_c^*}{dV^*}. \tag{34}$$

Since $\frac{dA_c^*}{dV^*} > 0$ and $\frac{dz_c^*}{dV^*} > 0$, this parameter combination results in a net predominance of the positive feedbacks, implying that the present-day Earth is located on the right side of the bifurcation diagram (Fig. 1). However, uncertain parameters impact the conclusion. If we for example take $\frac{dv^*}{dA_c^*} = \frac{4}{3}$ and $\frac{dv^*}{dz_c^*} = 0$ (for $\chi = 0$) instead, we get

$$\frac{d}{dV^*}\left(\frac{\dot{V}_{net}}{\dot{V}_{prod,E}}\right) = 0.36\frac{dA_c^*}{dV^*} - 0.2\frac{dz_c^*}{dV^*}. \tag{35}$$

In this case, the net growth rate increases with area but decreases with thickness. Here, the change of the continental area and thickness with the volume determines the combined strengths of the feedbacks, with positive feedbacks dominating for $\frac{dA_c^*}{dV^*} > \frac{5}{9}\frac{dz_c^*}{dV^*}$.

### 4. Earth's Recent History

Anytime in the evolution, there is least one specific value for the continental volume for which the continental production rate would equal the continental erosion rate. This value for the continental volume represents the fixed point. If negative feedbacks dominate over positive feedbacks, the fixed point is stable. In contrast, dominating positive feedbacks would imply that the fixed point is unstable. The volume of the continental crust with respect to the fixed point at any time in the evolution determines the net growth rate of continental crust. If the continental volume is in the vicinity of a stable fixed point it will evolve towards the fixed point. In contrast, if the continental volume is in the vicinity of the unstable fixed point, it will evolve away from it. The (absolute) net growth rate increases with the distance to the fixed point, and the net growth rate directly at the fixed point is zero. In Section 4.1, we discuss how fixed points evolve with time for scenarios with dominating positive and negative feedbacks and show the implications for the continental growth curves. In Section 4.2, we quantitatively analyze the feasibility of remaining in a state of small positive net continental growth during certain time intervals in the evolution for our model with dominating positive feedbacks and for a simplified model with dominating negative feedbacks.

### 4.1 Movements of the Fixed Points and Feasible Continental Growth Curves

The rates of continental production and erosion decrease with time, but not to the same extent. In our model, time-dependence is caused by the decreasing plate speed. Continental production depends more strongly on the plate speed than continental erosion does (which includes mechanisms not directly related to the plate speed, such as surface erosion). Therefore, the net growth rate for a given continental volume decreases with time.

However, the net growth rate is not only a function of time but also of the continental volume itself. To maintain a zero net growth rate (i.e. a fixed point), a decrease in mantle temperature must be compensated by a change of the continental volume. Therefore, an unstable fixed point (implying an increasing net growth rate with continental volume) moves to larger values of continental volume with time. A stable fixed point, on the contrary, moves to smaller values of continental volume.

In Fig. 3, we qualitatively illustrate the movements of the fixed points with mantle cooling and possible corresponding continental growth curves. We compare possible evolutions for a scenario with dominating negative (Fig. 3a) and positive (Fig. 3b) feedbacks. The movements of stable fixed points and of the unstable fixed point are represented by solid and dashed lines, respectively. Since our analysis is valid for the present-day tectonic regime only, we leave the early evolution open. Fig. 3a illustrates the general problem of thermal evolution models with continental growth with dominating negative feedbacks: here, two continental growth scenarios are feasible. One possibility is a negative present-day net growth rate (illustrated with paths 1 and 2). This would be the case if equilibrium between continental production and erosion has been reached in the past and the continental volume evolves in equilibrium following the evolution of the stable fixed point. However, this scenario is not consistent with most geochemical models (e.g., Stein and Ben-Avraham, 2007; Belousova et al., 2010; Dhuime et al., 2012). Alternatively, production and erosion rates may have been far away from equilibrium for a long time and may have only recently approached the fixed point (path 3). However, this would make a small net growth rate over billions of years difficult to explain.

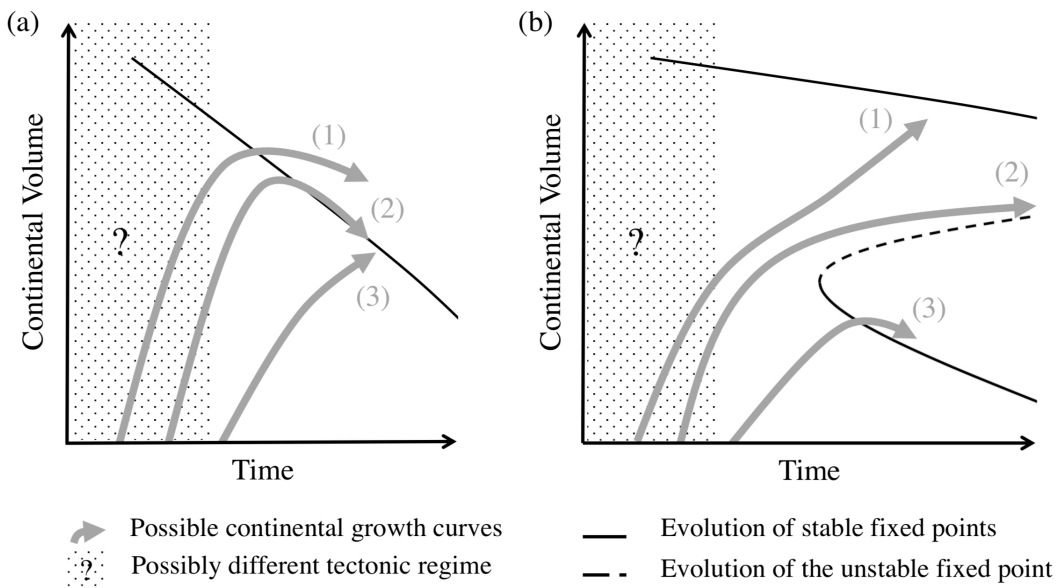

Fig. 3: Qualitative evolutions of equilibrium points of the continental volume (black lines; solid: stable, dashed: unstable) and corresponding continental growth curves (grey lines). The dotted area represents Earth's early history where continental production mechanisms where presumably different from today. (a) Model featuring only one fixed point, which is stable. This model results in either larger continents in Earth's history and a negative growth rate (paths 1 and 2), or a rather large present-day growth rate with the equilibrium not being reached yet (path 3). (b) Scenario of multiple fixed points with the intermediate being unstable. The location of the present-day Earth near the unstable fixed point allows a small present-day net growth rate (path 2), but would imply that depending on initial conditions, also evolutions towards the upper (path 1) and lower (path 3) fixed point are possible.

Instead, if positive feedbacks cause the emergence of an unstable fixed point at a certain time in the evolution, the continental growth trajectory may have already reached larger values than that corresponding to the newly emerged unstable fixed point (Fig. 3b, path 2). A small distance between the unstable fixed point and the continental growth trajectory results in a small net growth rate. The continental growth trajectory moves away from the unstable fixed point, that is, towards larger values of the continental volume. At the same time, the unstable fixed point moves to larger values as well, thereby sustaining a small distance to the continental growth trajectory and a small net growth rate. As a result, the proposed model allows the continental volume of the Earth to remain in a state of small positive net growth for a longer period of time.

Note that dominating positive feedbacks imply a critical dependence of the continental growth curve on initial conditions: Paths 1, 2 and 3 may start with similar initial conditions but evolve towards very different fixed points, however. In Section 4.2, we quantitatively analyze the plausibility of these scenarios and the range of initial values that result in small net growth rates.

### 4.2. Feasibility of a Small Net Growth Rate and Sensitivity to Initial Conditions

The present-day net continental growth rate is commonly thought to be significantly smaller than the rates in the early evolution of the Earth. Hawkesworth et al. (2018) argue that 3 billion years ago the continental volume comprised already 65-70% of the present value. Similarly, Condie (2014) estimates that 70% of the present-day continental volume was formed before 2.5 billion years ago. Dhuime et al. (2017) argue for a small average net growth rate between 0.6 and 0.9 $km^3 yr^{-1}$ during the past three billion years, and Armstrong (1991) even argues for an early rapid growth of continental crust and a steady state (i.e. zero net growth) since then. In this section, we combine our analytical models of continental production and erosion with a simple planetary evolution model to address the question of how long such a small net growth rate can be maintained for dominating negative and positive feedbacks.

For simplicity, we keep the continental thickness constant in this analysis, setting $\dot{V}_{net} = \dot{A}_{net} \frac{V_E}{A_E}$. From Eq. (10), we get

$$\frac{\dot{A}_{net}}{A_E} = \frac{\dot{V}_{prod,E}}{V_E} \left[ (v^* - 1) - (\phi_{surf} + \phi_{dela})(A_c^* - 1) - \phi_{subd}(v^* L_S^* - 1) \right], \qquad (36)$$

with $\dot{V}_{prod,E} = 5.25 \frac{km^3}{yr}$ (Stern, 2011) and $V_E = 7.18 \cdot 10^9 \ km^3$ (Schubert and Sandwell, 1989). Note that the continental production rate is calculated independently of the total length of the ocean-continent subduction zones, assuming that continental production along ocean-ocean subduction zones will compensate for a smaller length of ocean-continent subduction zones. Calculating the subduction erosion rate, we first keep the total length of ocean-continent subduction zones constant, i.e. $L_S^* = 1$. Later we will show that calculating $L_S^*$ following Eq. (7) does not change our results significantly.

For the evolution model, we assume that the total surface heat flow is proportional to the heat production rate, which is a reasonable approximation for the recent past, where the Urey ratio remained approximately constant (e.g., Jaupart et al., 2007). Hence, we multiply the heat flow (Eq. 16) with the relative heat production rate, which gives for our reference model

$$v^*(A, t) = \left( \frac{Q_{prod}(t)}{Q_{prod,E}} \right)^2 \left( \frac{1-\alpha}{1-\alpha A^*} \right)^2, \qquad (37)$$

where $\frac{Q_{prod}(t)}{Q_{prod,E}}$ is the heat production rate relative to the present-day rate. Calculating $\frac{Q_{prod}(t)}{Q_{prod,E}}$, we use half-lives and heat production rates of radiogenic elements following Korenaga (2008), combined with element abundances following Salters and Stracke (2004).

We compare this model with a model with dominating negative feedbacks, where we neglect the dependence of the plate speed on continental area, simply setting $v(t) \propto Q_{prod}(t)^2$. Here, the negative feedbacks related to continental erosion dominate and the fixed point is stable.

In Fig. 4, we show the net continental growth rate in a phase plane spanned by the relative heat production rate and the relative continental volume. We compare the model that neglects the positive feedback (Fig. 4a) with our reference model where the plate speed depends on the continental volume (Fig. 4b). Converting the time into the relative heat production rate, we also plot evolution trajectories for both models. The initial continental volumes at 1.5 Gyr are chosen such that the evolution trajectories exactly match the present-day continental volume at 4.5 Gyr with a zero net growth rate. Note that these values of the initial continental volume allow for the smallest possible continental growth rates that never fall below zero.

Both models can explain a small positive net growth rate during the past 1 Gyr. However, going back to the past 2 or 3 billion years, the model with strong positive feedbacks (Fig. 4b) allows for a much smaller net growth rate than the model without the positive feedback (Fig. 4a). The reason is that for smaller continental volume in the Earth's history, positive feedbacks reduce the net growth rate, thereby counteracting the hotter mantle and the larger heat production rate.

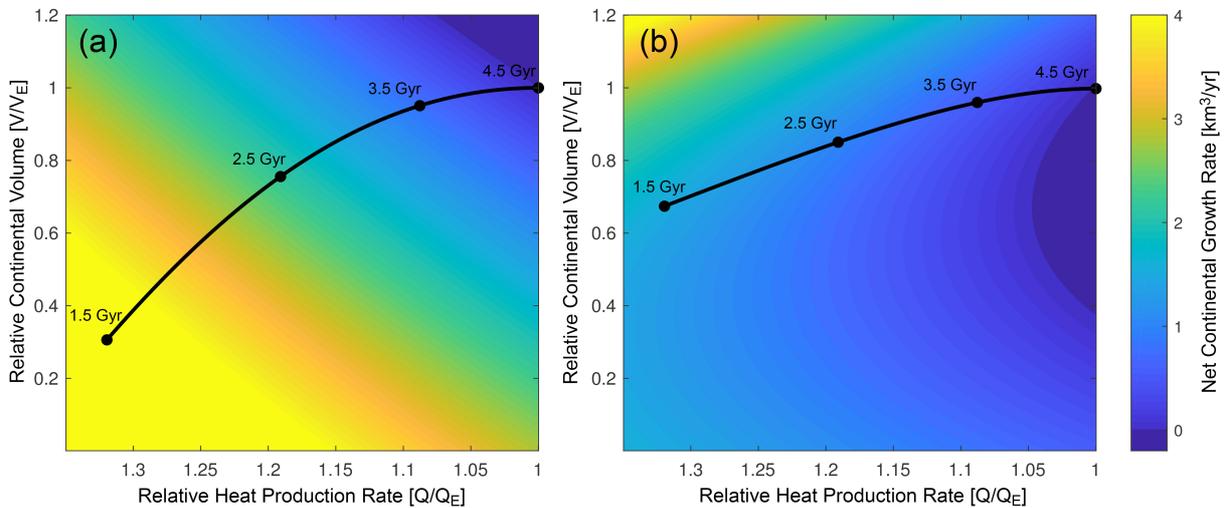

**Fig. 4:** Net continental growth rate as a function of the relative continental volume and the relative heat production rate. In Fig. 4a, the positive feedback that relates the plate speed to the continental volume is neglected, and the heat flux per unit area is set proportional to the heat production rate. Fig. 4b accounts for the positive feedback, and the heat flux per unit area is set proportional to the heat production rate divided by the surface area that is not covered by continents. The purple areas indicate negative net growth rates. Converting the time into the relative heat production rate, we plot evolution trajectories (black) with initial conditions that are chosen such that the net growth rate at 4.5 Gyr is zero.

Having the present-day Earth exactly located at a fixed point, such as in Fig. 4, would require precisely matching initial conditions. A larger initial continental volume for the model presented in Fig. 4a would cause the trajectory to evolve into the purple area implying negative net growth, which we exclude. On the contrary, a smaller initial continental volume would result in a positive present-day (and a larger average) net growth rate. The opposite is true for Fig. 4b, where a larger initial continental volume would result in a positive present-day (and larger average) net growth rate, whereas a smaller initial continental volume would imply negative net growth at 4.5 Gyr.

Exactly matching initial conditions, such as chosen for the model presented in Fig. 4, is unlikely for the real Earth system. In the following, we explore the effect of a difference to these initial values on the average net growth rate. We will term the initial continental volume $V_o$ at time $t_o$ and the present-day continental volume $V_1$ at time $t_1$. In Fig. 5, we plot the average net growth rate (left ordinate axis) and the total net amount of growth (right ordinate axis) during the time interval from $t_0$ to $t_1 = 4.5 \, Gyr$ with (a) $t_0 = 3 \, Gyr$, (b) $t_0 = 2.5 \, Gyr$, (c) $t_0 = 2 \, Gyr$, (d) $t_0 = 1.5 \, Gyr$. We compare the model with dominating positive feedbacks (blue) with the model with dominating negative feedbacks (red). The solid and the dashed lines are plotted for $L_s^* = 1$ and for $L_s^*(A_c^*)$ following Eq. (7), respectively. Note that a variable subduction zone length does not change the results significantly. The abscissa represents the (absolute) distance in units of $V/V_E$ ($V_E$ being the continental volume of the present-day Earth) of the initial continental volume to the limit value of $V_0 \left( \frac{dV_1}{dt} = 0 \right)$ that would result in a present-day zero net growth rate. The smallest possible average net growth rate for each scenario is available for a distance of zero and the average net growth rate and the total net growth increase with the distance from the limit value.

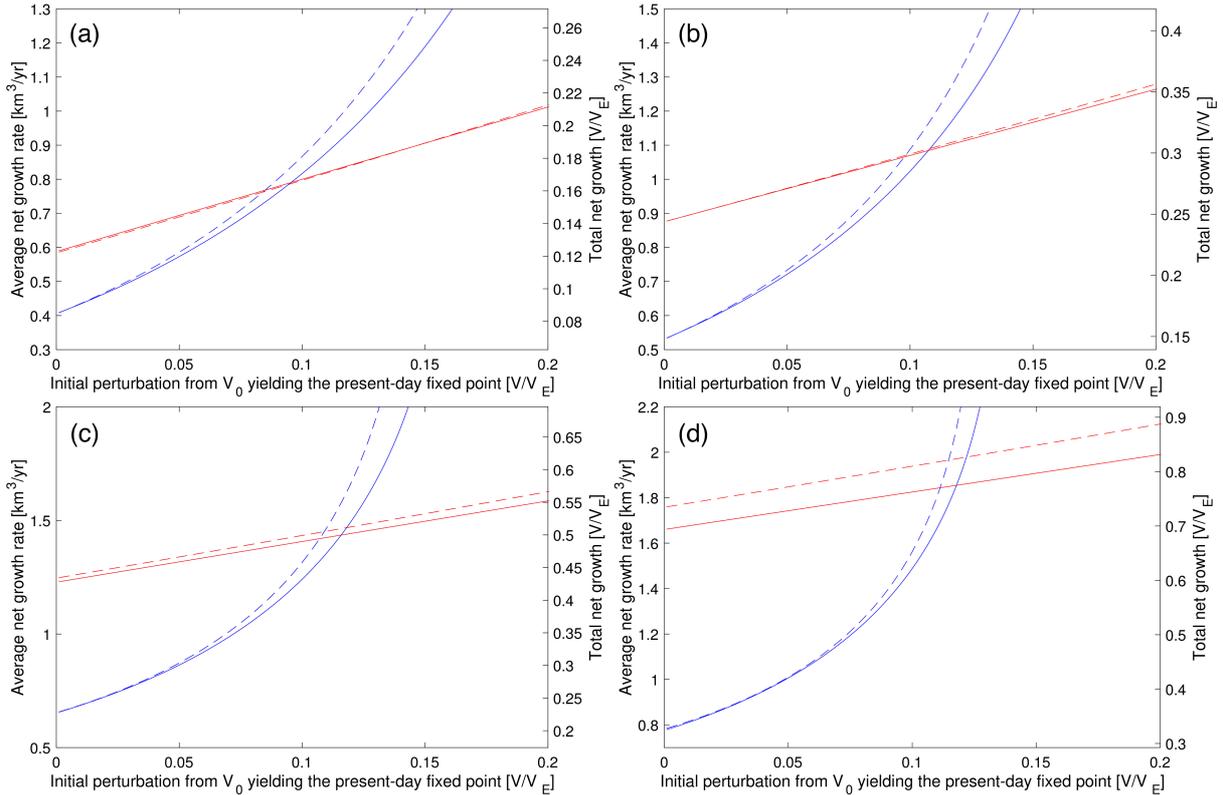

**Fig. 5:** Average net growth rate (left y-axis) and total net growth (right y-axis) for the model with dominating positive feedbacks as presented in this paper (blue) and for a model with dominating negative feedbacks (red). The dashed lines account for a variable subduction zone length while the solid lines are plotted for a constant subduction zone length. Fig. 5a, b, c, and d accounts for the last 1.5, 2.0, 2.5, and 3.0 billion years, respectively. The minimal feasible net growth rate during these periods for each model is plotted at a distance of zero at the x-axis. This rate results in the continental growth trajectory reaching the fixed point at 4.5 billion years and requires a precisely matching continental volume at 1.5 (or 2.0, 2.5, 3.0) billion years ago. With increasing distance on the x-axis, the average net growth rate increases. For a small distance on the x-axis (smaller than 10% of the present-day continental volume), the model with dominating positive feedbacks can explain a smaller average net growth rate than a model with dominating negative feedbacks.

For the model with dominating negative feedbacks (red), the initial continental volume $V_0$ is of minor importance and the slopes of the red lines in the Figure are small. This is so because the stable fixed point attracts the continental growth trajectories from a wide range of initial values. For the model with dominating positive feedbacks (blue), the initial conditions are important, however, and the average growth rate and the total net growth increase rapidly with distance from the limit value. This is so because the continental volume would evolve away from the unstable fixed point if the initial continental volume is too large or too small. However, if $V_0$ is within a small distance to a specific value, the average net growth rate will be smaller than any average net growth rate obtained from a model with dominating negative feedbacks (red curve). This distance can be read as the value of the abscissa of the point where the red and blue curves cross.

Our conclusion depends on the time interval for which a small net growth rate should be explained. Fig. 5a is plotted for the past 1.5 billion years. Here, in both scenarios, small net growth rates, between 0.6 and 0.9 $\frac{km^3}{yr}$ (as suggested by Dhuime et al., 2017) are feasible for a comparatively large range of initial continental volumes. However, a small average net growth rate during longer periods of Earth's history is more difficult to obtain. An average net growth rate between 0.6 and 0.9 $\frac{km^3}{yr}$ for the past 2 billion years (Fig. 5b) is feasible only using the model with dominating positive feedbacks, where an initial perturbation (applied to $V_0$) of up to 10% $V_E$ can be accepted. Such a small average net growth rate can also be maintained since the late Archean, i.e. for the past 2.5 billion years (Fig. 5c), using the model with dominating positive feedbacks. For the model with dominating negative feedbacks (red curve), the smallest possible net growth rate during this time period is 1.2 $\frac{km^3}{yr}$. This is even more true if an epoch of 3 billion years is considered for which the smallest growth rate with dominating negative feedbacks is 1.65 $\frac{km^3}{yr}$. For the model with dominating positive feedbacks (Fig. 5d, blue curve) the initial volume must be approximately 0.65 $V_E$ allowing only a small perturbation to be consistent with geochemical models that require an average net continental growth of 35% or less during this time interval as suggested by Hawkesworth et al. (2018).

Altogether, a small positive continental net growth rate during major periods of the Earth's history favors a model with dominating positive feedbacks in continental growth. This finding has implications on the continental volume expected on extrasolar Earth-like planets. Contrary to continental growth models with dominating negative feedbacks, planets with different initial conditions would result in a broad range of possible continental growth curves. If the perturbation (distance on the x-axis in Fig. 5) of the initial continental volume at $t_0$ exceeds 10 or 15% $V_E$, positive feedbacks would cause a runaway of the continental volume (blue curves). One would then expect a majority of extrasolar planets covered by a much smaller or larger fraction of continental crust.

### 5. Discussion

In this paper, we illustrate the following problem: A cooling mantle should reduce the continental production rate. The global heat flow decreases, thereby affecting the convection and subduction rate, and a decreasing mantle temperature reduces the melt fraction in subduction zones. If the effect of mantle cooling on the continental erosion rate is smaller, the stable equilibrium between continental production and erosion (i.e., the stable fixed point) moves to smaller values of the continental volume as the mantle cools. This in turn makes a small positive net growth rate over an extended time interval, which is suggested by geochemical studies, difficult to explain.

We propose strong positive feedbacks in the continental growth system, possibly dominating over negative feedbacks, as a solution to this problem. In this case, the continental production rate does not necessarily decrease with time, since positive feedbacks related to an increasing continental size counteract the effects of mantle cooling. A simple stability analysis based on linearized feedbacks indicates that such a scenario is plausible. However, there are large uncertainties related to the positive feedback that is considered in our paper, which we discuss in Section 5.1. On the contrary, further positive feedbacks that are not included in our stability analysis are discussed in Section 5.2. Possible negative feedbacks, which may emerge through a coupling of Earth's biosphere to the geology, are not included in the model and discussed in Section 5.3.

Another possible solution to the sketched problem could be a continental production rate that is hardly affected by the thermal state of the mantle. This possible solution is discussed Section 5.4. In Section 5.5, we discuss a further possible solution, that is a continental erosion rate that strongly depends on the thermal state of the mantle.

### 5.1 Continents, Heat Flux, and Plate Speed

In estimating the strength of the positive feedback, we use several crucial simplifications. First, we assume that the total mantle heat flow does not depend on the volume of the continents. Since continents are widely agreed to insulate the mantle, this implies that most of the heat finds its way through the oceanic crust, causing an increasing oceanic heat flux per unit area with continental coverage. Laboratory experiments and numerical simulations with temperature-dependent viscosity by Lenardic et al. (2005) indicate that for a fraction of the continental area up to the present-day Earth value of 0.4, continents may redistribute the heat flow towards the oceanic crust. However, if continents reduce the global heat flow instead, for instance by moving heat sources from the mantle to the continental crust or by reducing the global cooling rate, the positive feedback would be weakened. Therefore, our approach provides an estimate of the maximum strength of the positive feedback. If the global mantle heat flow is significantly reduced with increasing continental volume, negative feedbacks could dominate. To address this issue, we set the global mantle heat flow relative to its present-day Earth value, $Q_m^*$ in Eq. (12), to scale with the area of the oceanic lithosphere, i.e.

$$Q_m^* = \left(\frac{A_E - A_c}{A_E - A_{c,E}}\right)^{\zeta},$$

(38)

where $0 \leq \zeta \leq 1$ is a parameter controlling the dependence of the global mantle heat flow on continental area. For $\zeta = 0$ the total heat flow does not depend on continents, such as assumed in Section 3, while for increasing $\zeta$ continents increasingly influence the global mantle heat flow. For a constant continental thickness, the mantle heat flow per unit area (compare Eq. 15) can be written as

$$q_{mo}^* = \left(\frac{1-\alpha}{1-\alpha A^*}\right)^{1-\zeta},$$

(39)

which gives for the plate speed (compare Eq. 16)

$$v^* = \left(\frac{1-\alpha}{1-\alpha A^*}\right)^{2-2\zeta}.$$

(40)

Differentiating with respect to the continental area and setting $A^* = 1$ and $\alpha = 0.4$ (compare Eq. 17) yields

$$\frac{dv^*}{dA^*} = \frac{4}{3}(1-\zeta).$$

(41)

We can now determine the combined strength of the feedbacks (compare Eq. 33) depending on $\zeta$ (still assuming constant continental thickness):

$$\frac{d}{dV^*}\left(\frac{\dot{V}_{net}}{\dot{V}_{prod,E}}\right) = \frac{dA_c^*}{dV^*}\left(\frac{4}{3}(1-\phi_{subd})(1-\zeta) - \phi_{surf} - \phi_{dela}\right) \qquad (42)$$

We can calculate a critical value $\zeta_{crit}$ for which $\frac{d}{dV^*}\left(\frac{\dot{V}_{net}}{\dot{V}_{prod,E}}\right) = 0$. Using the parameter values of Tab. 1, we get $\zeta_{crit} = 0.4$. For $\zeta < \zeta_{crit}$, positive feedbacks dominate, with $\zeta_{crit}$ being the critical value for which a bifurcation occurs in the continental growth system.

Note that if continents reduce the global mantle cooling rate significantly, this may also cause a sensitivity of continental growth to initial conditions: A larger continental volume compared to a reference case would then cause a larger mantle temperature in the subsequent evolution compared to this refence case. With continental production dependent on the mantle temperature, this difference in the continental volume may be amplified as the planet evolves. Similar to dominating positive feedbacks, Earth-like planets would then be expected to show a large diversity in their continental volume. However, further work is required to analyze this sensitivity in more detail.

### 5.2 Effects of sediments as additional positive feedbacks

Positive feedbacks related to subducted sediments have been neglected in our analysis. The sedimentation rate increases with the surface area of continental crust. Although sediments also reach passive margins, the rate at which sediments enter subduction zones increases with the sedimentation rate. The most obvious positive feedback is that production of new continental crust in subduction zones requires the presence of water (Campbell and Taylor, 1983), which is partly subducted in stable hydrous phases within sediments. When released in the source region of partial melt, water lowers the melting temperature of the subducting oceanic crust and mantle material. Partial melt with an andesitic composition is generated and migrates upwards, eventually causing the formation of new continental crust. The contribution of sediments to water subduction varies greatly among subduction zones, depending on, e.g., their temperature profile (Ono, 1998). Although subducting sediments contain roughly as much water as the oceanic crust when entering a subduction zone (Jarrard, 2003), significant quantities of pore- and loosely-bound water are expelled at shallow depth. However, even at post-arc depths, subducted sediments can still contain $\approx 5 \, wt\%$ of the water, e.g. stored in clay-rich layers (Hacker, 2008). Low-temperature metamorphic reactions (e.g., Stern, 2002) may form hydrous minerals, which possibly stabilize water during subduction transporting it to greater depth. Furthermore, hydration of forearc-mantle material that is advected further down by induced convection could also enhance water subduction (Deschamps et al., 2012; 2013).

The effect of water carried by sediments into the source region of partial melt on continental production depends on the distribution of the water in the melt zone. It is not clear whether additional water would increase the melt fraction while not enlarging the melt zone spatially, or whether additional sediment water would rather enlarge the melt zone spatially while keeping the melt fraction constant. In both cases, additional water by subducted sediments can increase the volume rate at which melt is created and thereby the production of continental crust. However, particularly in the first scenario, the magnitude of this effect is difficult to constrain due to the nonlinear effect of water on the melting temperature (e.g., Katz et al., 2003).

In addition to the subduction of water in sediments, a large variety of effects of sediments on continental production has been described, which we summarize here. First, subducted sediments enrich the source region of partial melt with incompatible elements, facilitating the generation of partial melt of andesitic composition (e.g., Gazel et al., 2015). Manning (1996) argues that the presence of sediments along the slab-mantle interface increases the aqueous silica concentration, eventually affecting the production and overall composition of continental crust. Rosing et al. (2006) point out that clay minerals produced during weathering can act as alkali exchange media, thereby also promoting the production of continental crust. Finally, an important role is played by the physical properties of the subducting sedimentary layer. Due to its low permeability, it may partially suppress dewatering at shallow depth (discussed in Höning et al., 2014). This would then result in a greater availability of water in the source region of partial melt, thereby enhancing the rate of continental production. In addition, the sedimentary layer insulates the subducting slab from the hot mantle, thereby stabilizing hydrous phases in the subducting oceanic lithosphere.

The quantification of these processes is complex and beyond the scope of this paper. However, we emphasize that these processes act as additional positive feedbacks, making the dominating positive feedbacks altogether more likely.

### 5.3 Volatile Cycles, Biology, and Long-Term Stabilization

The interior evolution of planets is closely coupled to their atmospheres (Foley, 2015; Foley and Driscoll, 2016; Lenardic et al., 2016b; Tosi et al., 2017), and Lee et al. (2008) argue that chemical weathering is an important process in producing felsic continents. Ultimately, the growth of continental crust is closely related to the evolution of the atmosphere, a link that is not addressed in this paper. An increasing mantle water concentration would imply a reduced surface water budget, which in turn would affect weathering rates, thereby possibly stabilizing the equilibrium point. The biosphere has been argued to play a role in weathering (e.g., Schwartzman and Volk, 1989), and may therefore also have an impact.

Life certainly benefits from the present-day continental configuration, with large emerged land and continental shelve areas facilitating harvesting solar energy by photosynthesis to a large extent, while at the same time large oceans promoting a wet climate. For the present-day continental coverage, subduction zones are presumably close to a maximum (Höning et al. 2014), going along with large continental shelf regions, which are areas of great phytoplankton primary production (e.g., Behrenfeld and Falkowski, 1997). If bioactivity in turn plays a role in the production of continental crust (Rosing et al., 2006: Höning et al., 2014), a stabilizing feedback may be established. Theories claiming that unusually rapid evolution of Earth's biosphere has been crucial to regulate Earth's system and to keep Earth habitable (Chopra and Lineweaver, 2016) may then be expanded: Later steps of biological evolution, for example the emergence of land plants and fungi strongly enhancing continental weathering, could have also been important if Earth's system developed a bistability in the more recent past.

A further possible long-term stabilization of the continental system it related to different timescales of the individual feedbacks. For simplicity, we considered all feedbacks to operate simultaneously when comparing the feedback strengths. This is certainly a simplification, since surface erosion would act as an instantaneous negative feedback for the continental volume, whereas feedbacks related to the plate speed require an adjustment of the Earth mantle system, which could imply a delay. If negative feedbacks stabilize the continental volume way faster than positive feedbacks act, the continental system could remain stable altogether.

### 5.4 Continental Production and Thermal Evolution

In our model, we set the rate of continental production proportional to the plate speed. This is reasonable, since the production of continental crust requires the supply of water in subduction zones, which is mainly stored in hydrous phases within the subducting oceanic crust. Using boundary layer theory, we scaled the plate speed with the heat flux via the Rayleigh number. Since mantle convection contributes to driving plate tectonics, a linear dependence between convection speed and plate speed is a reasonable first-order approximation. As a result, a cooling mantle reduces the rate of continental production, and the stable fixed point moves to smaller values of the continental volume with time, making a small positive net growth rate over a large time interval challenging to explain. However, a reducing effect of mantle cooling on the rate of continental production is not inevitable.

Continental growth mechanisms in the Archean may have been different then today (e.g., Rudnick, 1995). The onset time of modern-style plate tectonics is discussed controversially (e.g., see review by Condie and Kröner, 2008). The production of continental crust in the Archean was possibly more widespread and not restricted to continental margins. Plume-induced recycling and melting (e.g., Rozel et al., 2017; Bédard, 2018) and the delamination of the lower mafic crust (Sizova et al., 2015) are considered feasible mechanisms to produce Archean continental crust without subduction. Gazel et al. (2015) propose melting of enriched oceanic crust in hot subduction zones as a mechanism of continental production during the Archean. A transition to the present-day tectonic regime has been argued to cause a phase of rapid continental growth (Taylor and McLennan, 1985; Labrosse and Jaupart, 2007; Dhuime et al., 2012; 2015). We restrict our present model of continental growth to the time after rapid growth extending back to 3 billion years at most. (Fig. 5d).

It has been argued that a hotter mantle in Earth's history would imply a thicker oceanic crust than today, which would less readily subduct (Sleep and Windley, 1982) resulting in more sluggish plate tectonics (Korenaga, 2006). Ultimately, large local mantle temperatures may cause a transition in tectonic regime (Lenardic et al., 2005). These processes could strongly reduce the decrease of the plate speed with time. If the plate speed increased with time rather than decreasing, it might be possible that the continental production rate increases as the mantle cools, which would imply that the fixed points move in opposite directions than discussed in Section 4. That is, the stable fixed points would move to increasingly larger values of the continental volume while the unstable fixed point would move to smaller values.

Our model assumes the validity of simple boundary layer theory when coupling the heat flux to the convection speed and plate speed. Semi-empirical thermal evolution models that use the concept of a maximum age of oceanic plates (Labrosse and Jaupart, 2007) result in a slower mantle cooling than classical boundary layer models (e.g., Stevenson et al., 1983; Höning and Spohn 2016). Geological evidence for the Archean mantle temperature comes from komatiites, ultramafic volcanic rocks formed during this period. The formation of komatiites requires comparatively large degrees of partial melting and thus either a mantle roughly 200-300 K hotter than today (Arndt et al., 1998) or the presence of water to sufficiently reduce the melting temperature. In the latter case, a mantle temperature about 100 K higher than today (Grove and Parman, 2004) would suffice. If the mantle temperature and the plate speed did not change significantly during the past 3 billion years, smaller average net continental growth rates than those discussed in Section 4.2 could be obtained.

### 5.5 Continental Erosion and Thermal Evolution

In our model, we assumed that subduction erosion depends linearly on the plate speed, whereas we neglected a dependence of surface erosion and lower crustal delamination on tectonics. As a result, the total continental erosion rate only weakly depends on the mantle temperature, and the stable fixed point moves to smaller values of the continental volume as the mantle cools. This is certainly a simplification: Surface erosion is particularly efficient in regions of steep topography (e.g., Flament et al., 2008; 2013), whereas the erosion rate in the flat interior continents is small (Blackburn et al., 2012). Steep topographies in turn are a result of orogeny, and therefore dependent on tectonics. Lower crustal delamination may depend on densification caused by metamorphic eclogitization (Krystopowicz and Currie, 2013), a process that mainly occurs at convergent plate boundaries and is therefore also related to tectonics. Whitehead (2017) argues that interactions between the ocean area, mountain building and erosion processes could contribute to form a well-balanced oceanic cistern, which may be stable in the long-term.

Altogether, a stronger dependence of the continental erosion rate on the plate speed, rather than on the actual continental volume, cannot be ruled out. This would imply that the continental erosion rate would strongly decrease as the mantle cools, similarly to what we model for the continental production rate. The stable fixed point would then not move rapidly to smaller values of the continental volume with mantle cooling, which would provide an alternative solution to the problem of explaining a small positive net continental growth rate over an extended time period.

### 6. Conclusions and Implications

If continental production depends on the thermal state of the mantle to a larger extent than continental erosion does, the stable equilibrium between continental production and erosion moves to smaller values of the continental volume as the mantle cools. We show that this inference makes the observed small positive continental net growth rate over a large time interval throughout Earth's evolution difficult to explain. The following solutions to this problem may be considered:

First, the continental production rate may only weakly depend on the thermal state of the mantle. This case is conceivable if a large mantle temperature causes a thick oceanic crust that may tend to resist subduction (Sleep and Windley, 1982) and cause sluggish plate tectonics (Korenaga, 2006). Second, the continental erosion rate may strongly depend on the thermal state of the mantle. A possible link is given by tectonic activity that causes on the one hand densification and crustal delamination, and on the other hand steep topographies that lead to large rates of mechanical erosion. However, further studies are required that explore how crucial the plate speed is for erosion to remain efficient, or whether the time required to erode regions of steep topography is the limiting factor. The third solution to the sketched problem is a net predominance of positive feedbacks in the continental growth system, which we explore in this paper.

We applied a stability analysis to the processes involved in continental growth with plate tectonics by identifying and analyzing positive and negative feedbacks. Our starting assumption is a present-day equilibrium between continental production and erosion. Depending on the modulus of the change of the net growth rate with continental volume $\frac{d\dot{V}_{net}}{dV}$, a bifurcation occurs with negative feedbacks dominating for $\frac{d\dot{V}_{net}}{dV} < 0$ and positive feedbacks for $\frac{d\dot{V}_{net}}{dV} > 0$ (compare Fig. 1). The former regime would be characterized by singular stable fixed point towards which the continental volume would evolve. The latter regime would have three fixed points of which the intermediate fixed point with the continental volume of the present-day Earth would be unstable. The two remaining fixed points would be stable and distinguished by smaller and larger values of the continental volume. We investigated the evolution of the fixed points for the two regimes to the left or the right of the bifurcation point and implications on continental growth curves.

Our results indicate that the strengths of positive and negative feedbacks in continental growth have a similar magnitude. Large uncertainties are due to uncertain mechanisms and parameter values such as the effect of continents on global mantle heat flow, the present-day value of subcontinental mantle heat flow, the coupling between speed of convection and plate speed, and on poorly quantified effects of subducted sediments. However, a first-order approximation indicates that a net predominance of positive feedbacks is plausible, implying that the present-day state of the continental volume is unstable.

Our model can explain the observed small average continental growth rate of approximately $0.8 \ km^3 yr^{-1}$ (e.g. Dhuime et al., 2017) applied to the past 1.5 billion years if the present-day continental volume is unstable, but also if negative feedbacks are dominating and the Earth would evolve towards a stable fixed point. But if the small growth rate is required to extend back to the end of the Archean, then the former situation is much better suited to satisfy the requirement: If continental production depends on the continental volume itself, additionally to the thermal state of the mantle, net continental growth in the Earth's history (for a smaller continental volume and a hotter mantle) is slower compared to a scenario where continental production does not depend on the continental volume.

In the scenario of dominating positive feedbacks continental growth strongly depends on initial conditions and the continental volume itself. For instance, a positive average net growth rate of $\leq 0.8 \ km^3 yr^{-1}$ for the past 1.5 billion years requires a certain continental volume 1.5 billion years ago or a difference to this value of up to 10% $V_E$ ($V_E$ is the present-day Earth's continental volume). Reproducing this rate over the past 2.5 billion years still allows an initial difference of 5% $V_E$. It is also possible to explain such a small average rate for the last 3 billion years with the scenario with dominating positive feedbacks, although this requires precisely matching initial conditions.

A large variation in the initial conditions in the scenario of dominating negative feedbacks is of minor importance. Here, varying the starting volume $V_0$ by 20% $V_E$ will change the average net growth rate by only 0.4 $km^3 \ yr^{-1}$. In contrast, in the scenario with dominating positive feedbacks, perturbations in the continental volume of more than 10% $V_E$ are significantly reinforced, causing the system to approach states of much smaller or larger continental volume. Earth-like exoplanets would then be expected to show a large variety in the continental volume. This indicates that bistability in planetary evolution (e.g., Sleep, 2000; Weller et al., 2015; Lenardic et al., 2016a) also includes the growth of continental crust, which has not been accounted for so far.

Models of the mantle water cycle that investigate the surface and mantle water budget for different planetary masses (e.g., Cowan and Abbot, 2014; Schaefer and Sasselov, 2015; Komacek and Abbot, 2016) are useful to give insights into the water-land ratio that could be expected on other planets. However, the surface area that has emerged above sea-level is also affected by the volume of continental crust. If positive feedbacks dominate the continental growth system, this has wide implications for the water-land ratio that could be expected on planets with plate tectonics beyond our solar system. Multi-branched functions (discussed by Sleep, 2000) could be applied to continental growth and cause rapid changes in the evolution of continental volume during the evolution of planets with plate tectonics. Depending on initial conditions, planets would then show a large diversity, ranging from almost the entire planet covered with dry continents up to water-worlds, even if they have similar ages, sizes, and total water inventories.

**Acknowledgement**

We thank Nicolas Flament, Norm Sleep, and two anonymous reviewers for their thoughtful reviews that greatly enhanced the quality of this paper. D. Höning and N. Tosi have been supported by the Helmholtz Association (project VH-NG-1017) and D. Höning has further been supported through the NWA StartImpuls.